\documentclass[a4paper,11pt]{article}
\usepackage{pos}

\usepackage{subfigure}

\usepackage{comment}

\usepackage[capitalize]{cleveref}
\Crefname{section}{Section}{Sections}
\Crefname{figure}{Figure}{Figures}
\Crefname{table}{Table}{Tables}
\Crefname{appendix}{Appendix}{Appendices}
\Crefname{equation}{Eq.}{Eqs.}
\newcommand{\Onlinecite}[1]{%
    \IfSubStr{#1}{,}{Refs}{Ref}.~\cite{#1}%
}

\DeclareMathOperator{\tr}{\rm tr\,}

\title{Investigating the flux tube structure within full QCD}

\author[a]{Marshall Baker}
\author[b]{Paolo Cea}
\author[c]{Volodymyr Chelnokov}
\author*[b]{Leonardo Cosmai}
\author[d,e]{Alessandro Papa}

\affiliation[a]{Department of Physics, University of Washington, 3910 15th Ave. NE, Seattle, WA 98195-1560, USA}

\affiliation[b]{INFN - Sezione di Bari, via Amendola 173,  I-70126 Bari, Italy}

\affiliation[c]{Institut f\"ur Theoretische Physik, Goethe Universit\"at, Max-von-Laue-Str. 160438 Frankfurt am Main, Germany}

\affiliation[d]{INFN - Gruppo collegato di Cosenza, I-87036 Arcavacata di Rende, Cosenza, Italy}

\affiliation[e]{Dipartimento di Fisica Universit\`a della Calabria, via Pietro Bucci, 87036 Arcavacata di Rende (CS), Italy}

\emailAdd{mbaker4@uw.edu}
\emailAdd{paolo.cea@ba.infn.it}
\emailAdd{chelnokov@itp.uni-frankfurt.de}
\emailAdd{leonardo.cosmai@ba.infn.it}
\emailAdd{alessandro.papa@fis.unical.it}

\abstract{A characteristic signature of quark confinement is the concentration of the chromoelectric field between a static quark–antiquark pair in a flux tube. 
Here we report on lattice measurements of field distributions on smeared Monte Carlo ensembles in QCD with (2+1) HISQ flavors. We measure the field distributions for several distances between static quark-antiquark sources, ranging from 0.6 fm up to the distance where the color string is expected to break.}

\FullConference{The 41st International Symposium on Lattice Field Theory (LATTICE2024)\\
 28 July - 3 August 2024\\
Liverpool, UK\\}

\begin{document}
\maketitle

\section{Introduction and lattice setup.}
Gaining a detailed understanding of color confinement remains a central goal of nonperturbative studies in QCD.
Lattice numerical simulations have long shown the formation of tube-like structures when examining the chromoelectric fields between static quarks.
The observation of these tube-like structures in lattice simulations is linked to the linear potential between static color charges and provides direct numerical evidence of color confinement.
In recent years, we have used numerical Monte Carlo simulations of SU(3) pure gauge theory on a spacetime lattice to investigate the detailed structure of the color fields around two static sources, a quark and an antiquark, 
at both zero~\cite{Baker:2018mhw,Baker:2019gsi,Baker:2022cwb} and nonzero temperatures~\cite{Baker:2023dnn}.

Our earlier investigations indicate that the  chromomagnetic field $\vec{B}$ around the sources is consistent with zero within the margin of uncertainty. 
The transverse components of the chromoelectric field $\vec{E}$, relative to the line connecting the sources, consist solely of a perturbative, irrotational, short-range contribution. 
The longitudinal component of  $\vec{E}$  can be decomposed into a perturbative, short-range part and a nonperturbative term,  $\vec{E}^{\text{NP}}$, which encodes the confining properties and takes the form of a smooth flux tube.

In this work, we explore the case of full QCD with dynamical quarks. We expect the results to hold in a manner similar to SU(3) pure gauge theory, as long as the distance between the static sources remains below the threshold where string breaking~\cite{Philipsen:1998de,Kratochvila:2002vm,Bali:2005fu,Koch:2015qxr} occurs and the flux-tube structure disappears. However, at the onset of string breaking, new features in the behavior of color fields and currents are expected to emerge, which can be detected in our numerical simulations.

As in our previous studies on SU(3) pure gauge theory~\cite{Baker:2018mhw,Baker:2019gsi,Baker:2022cwb}, we determine the spatial distributions of the color fields induced by a static quark-antiquark pair using lattice measurements of the connected correlation function  $\rho^\text{conn}_{W, \mu \nu}$ ~\cite{DiGiacomo:1989yp}, which relates to the plaquette  $U_P = U_{\mu \nu} (x)$  in the  $\mu \nu$  plane and a square Wilson loop  $W$:
\begin{equation}
    \rho^\text{conn}_{W, \mu \nu} = \frac {\langle\tr (WLU_PL^\dagger)\rangle}{\langle\tr(W)\rangle} - \frac{1}{N} \frac {\langle\tr (U_P) \tr (W)\rangle}{\langle\tr(W)\rangle}\;,
    \label{connected1}
\end{equation}
$N=3$ being the number of QCD colors. The correlator $\rho^\text{conn}_{W, \mu \nu}$ provides a lattice definition of a gauge-invariant field strength tensor $\langle F_{\mu \nu}\rangle_{q \bar{q}} \equiv F_{\mu \nu}$:
\begin{equation}
\rho^\text{conn}_{W, \mu \nu} \equiv~~ a^2 g\langle F_{\mu \nu}\rangle_{q \bar{q}} ~~ \equiv~~ a^2 g ~F_{\mu \nu}\;.
\label{connected2}
\end{equation}
\begin {figure}[htb]
\centering
\includegraphics[width=0.45\linewidth,clip]{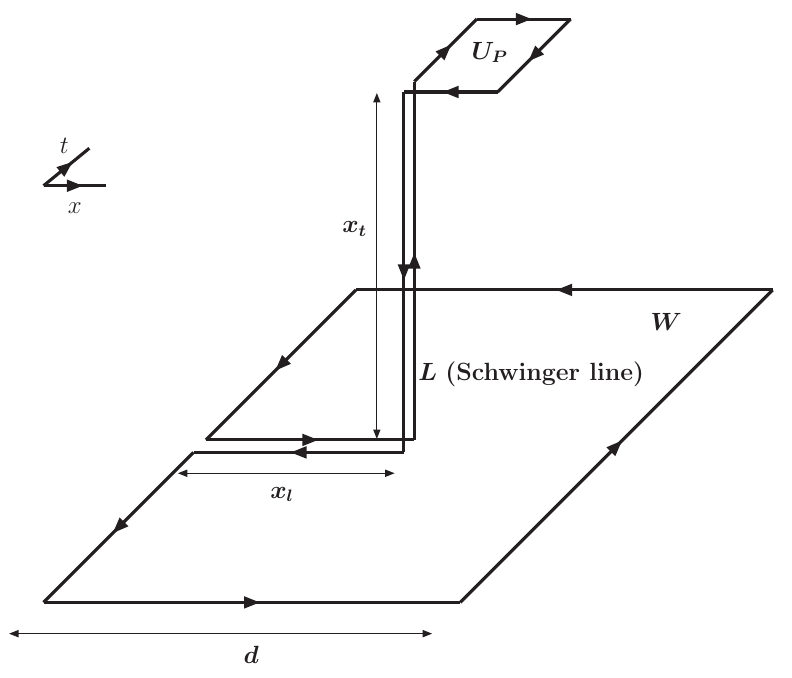}
\includegraphics[width=0.25\textwidth,clip]{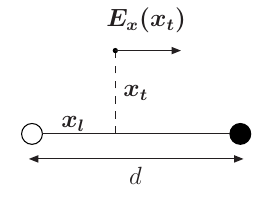}
\caption{The connected correlator between the plaquette $U_{P}$ and the Wilson loop (subtraction in $\rho_{W,\,\mu\nu}^\text{conn}$ not explicitly drawn).
The longitudinal electric field $E_x(x_t)$ at some fixed displacement $x_l$ along the axis connecting the static sources (represented by the white and black circles), for a given value of the transverse distance $x_t$.}               
\label{fig:op_W}
\end{figure}
\begin {figure}[thb]
\centering
\includegraphics[width=0.55\linewidth,clip]{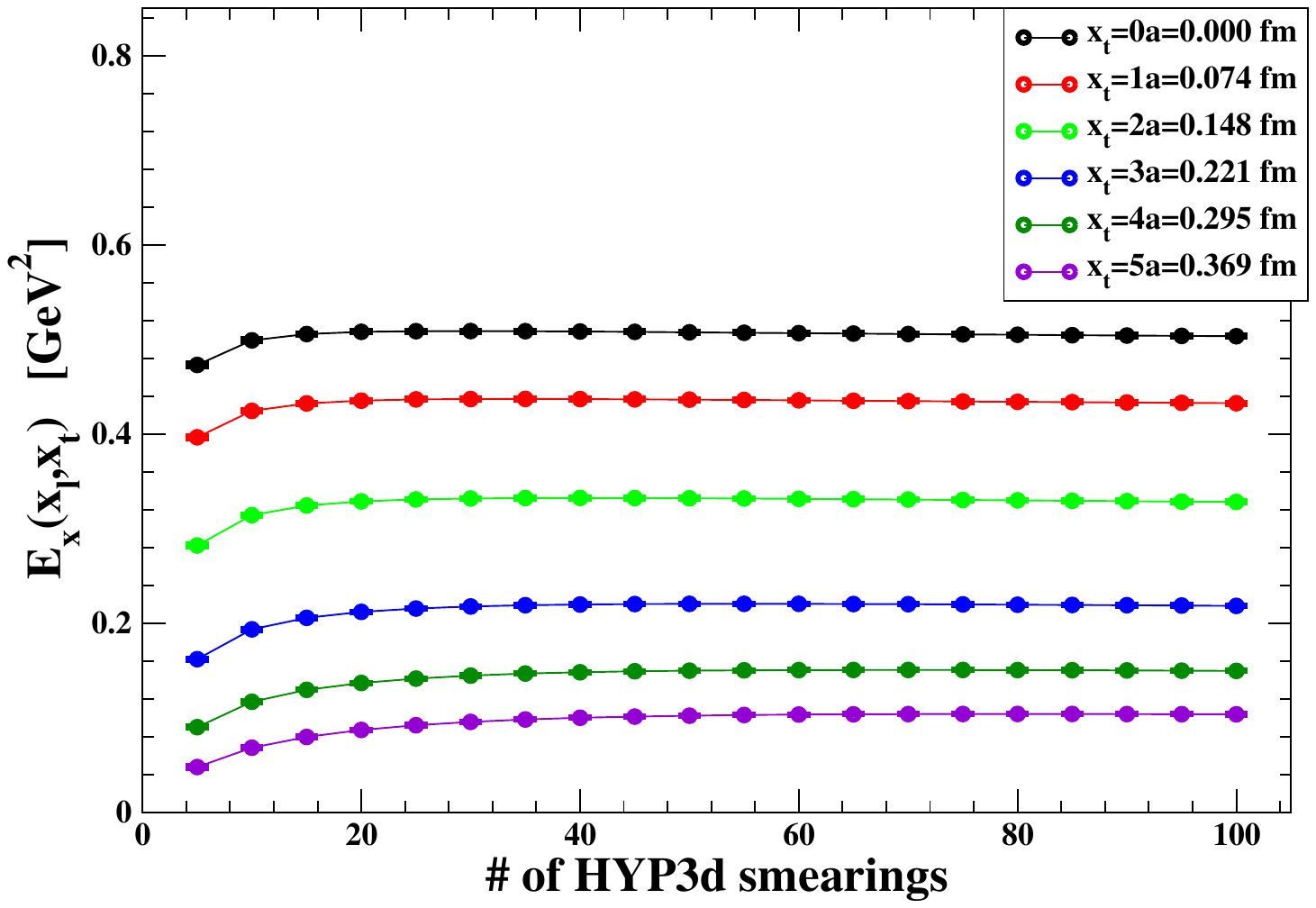}
\caption{Behavior under smearing of the "full" longitudinal electric field, $E_x$, for different values of the transverse distance $x_t$, at $\beta=7.158$ and $d=10a=0.74$ fm on a lattice $32^4$.}               
\label{fig:example_of_smearing}
\end{figure}
We perform lattice QCD simulations with 2+1 flavors of HISQ (Highly Improved Staggered Quarks)~\cite{Follana:2006rc, Bazavov:2009bb, Bazavov:2010ru}. 
The couplings are adjusted to follow a line of constant physics (LCP), as determined in Ref.~\cite{Bazavov:2011nk}, with the strange quark mass  $m_s$  fixed at its physical value and a light-to-strange mass ratio  $m_l/m_s = 1/20$, corresponding to a pion mass of 160 MeV in the continuum limit. 
We have simulated the theory for several values of the gauge coupling, using lattices of size  $24^4$,  $32^4$, and  $48^4$. 
The thermalized lattice configurations  are separated by 25 trajectories of rational hybrid Monte Carlo (RHMC) of unit length. 
The chromoelectromagnetic field tensor is measured for a static quark-antiquark pair placed at a fixed distance  $d$  in lattice units. 
The lattice spacing is determined through the observable  $r_1$, as defined in Appendix B of Ref.~\cite{Bazavov:2011nk}.
\begin{figure*}[htb]
\centering
\includegraphics[width=0.5\linewidth,clip]{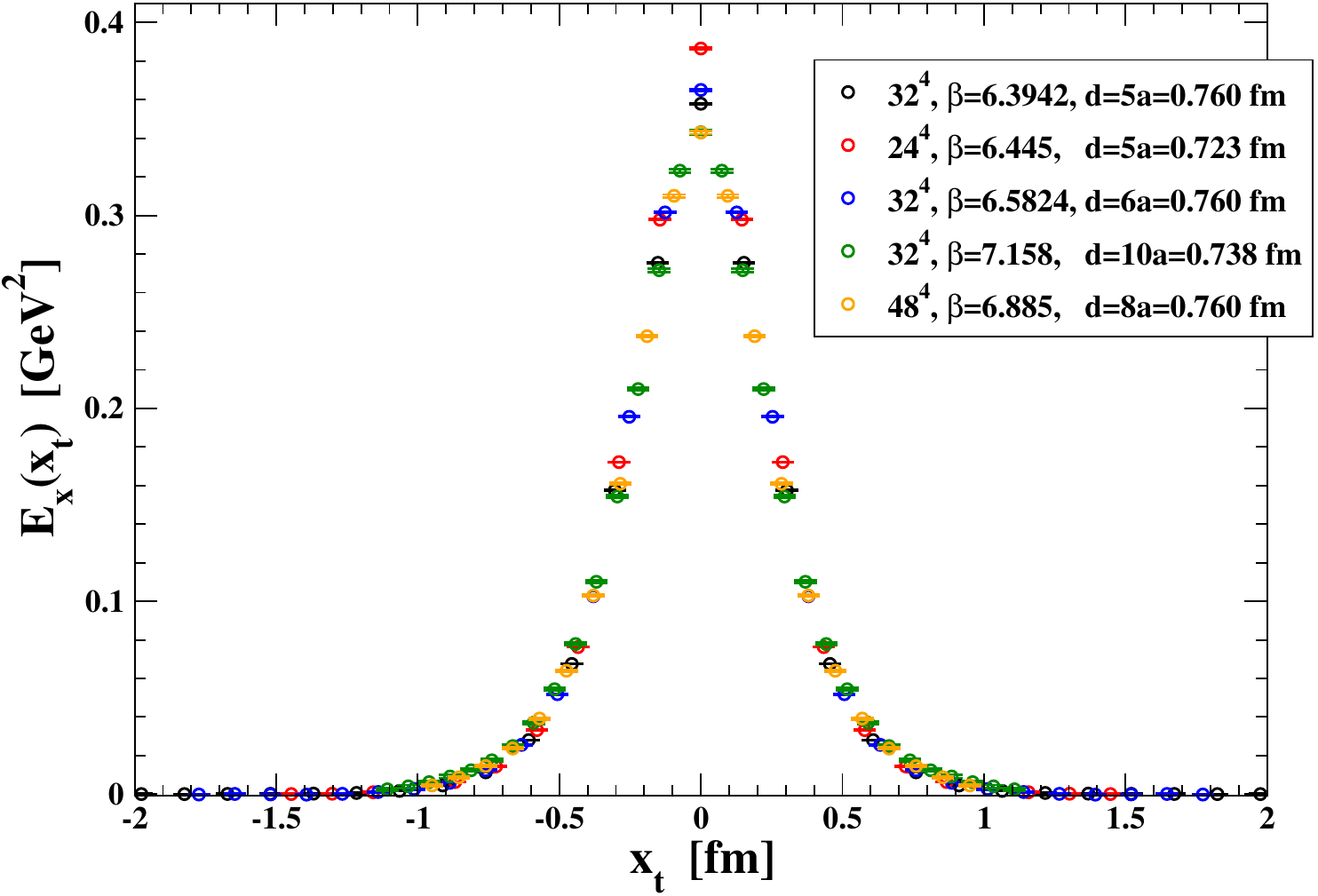}
\includegraphics[width=0.5\linewidth,clip]{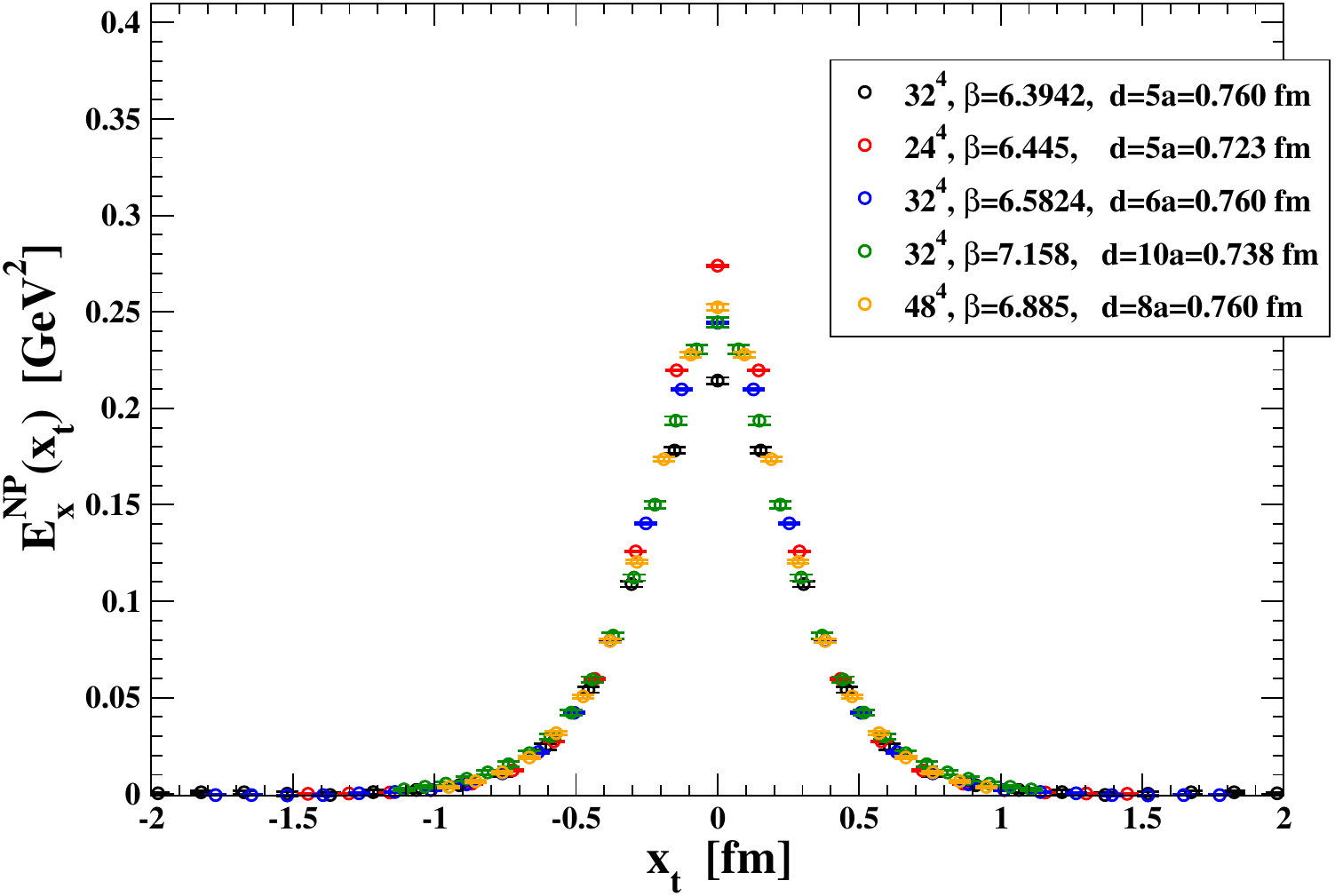}
\caption{Scaling test for the full electric field (top) and for its nonperturbative part (bottom) on the midplane for distances in the range $d=0.723-0.760$ fm.
}            
\label{fig:scaling}
\end{figure*}
\begin{figure*}[htb]
\vspace{-0.0cm}
\centering
\includegraphics[width=0.48\linewidth,clip]{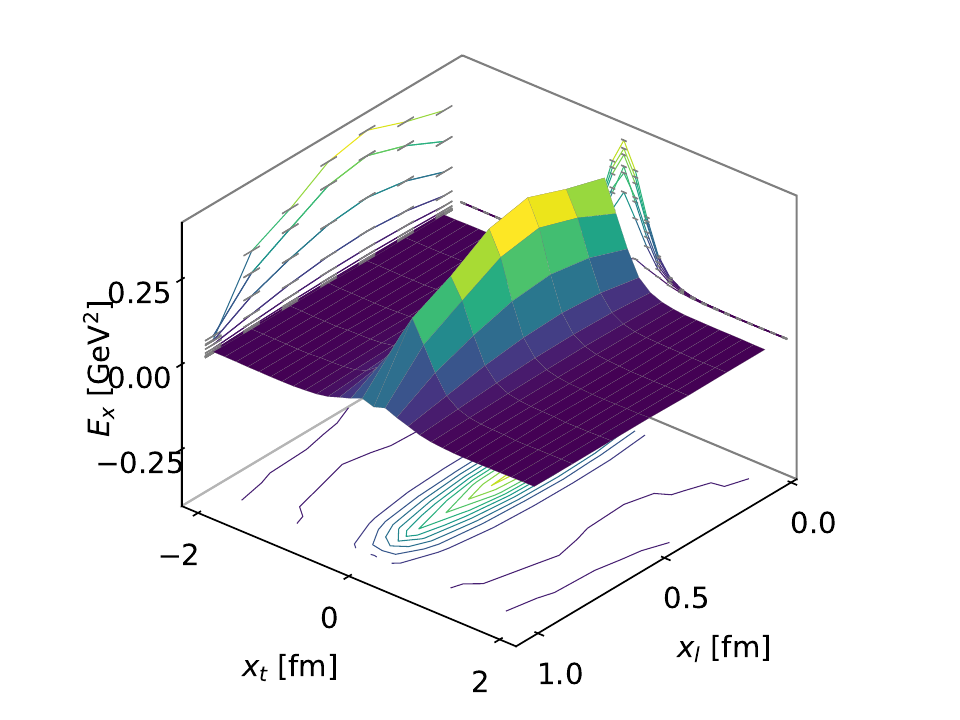}
\includegraphics[width=0.48\linewidth,clip]{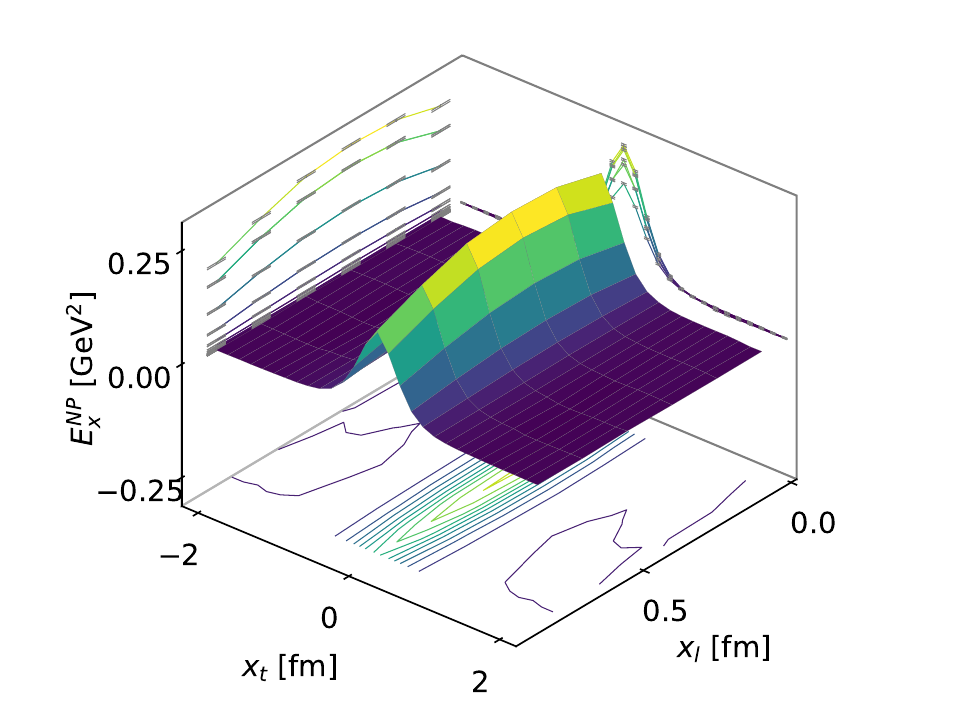}
\caption{Space distribution of the full electric field $E_x$ (left) and of its
nonperturbative part $E_x^{NP}$ (right) for $\beta=6.3942$ and $d=6a=0.91$ fm.}          
\label{fig:Ex_3d}
\end{figure*}
Since the connected correlator defined in Eq.~\ref{connected1} exhibits large
fluctuations at the scale of the lattice spacing, to improve the signal-to-noise ratio,
we smoothed out configurations by a {\em smearing} procedure. Moreover, the smearing behaves as an effective renormalization for the operator defined in Eq.~\ref{connected1}.
We apply one step of 4-dimensional hypercubic smearing~\cite{Hasenfratz:2001hp} on the temporal links (HYPt), with smearing parameters $(\alpha_1,\alpha_2,\alpha_3) = (1.0, 1.0, 0.5)$, and $N_{\rm HYP3d}$ steps of hypercubic smearing  restricted to the three spatial directions (HYP3d) with $(\alpha_1^{\text{HYP3d}},\alpha_3^{\text{HYP3d}}) = (0.75, 0.3)$.
We choose therefore the {\em optimal} number of smearing steps as the one for which the field takes its maximum value. It typically corresponds to a few units at small values of $x_t$ and to a few tens at larger values of $x_t$.

\section{Numerical results}

We have considered source distances over a wide range, from approximately 0.57 fm to 1.37 fm, corresponding to  $\beta$-values between 6.25765 and 7.158. 
To ensure our lattice setup is sufficiently close to the continuum limit, we verified that different choices of lattice parameters, corresponding 
to the same physical distance  $d$  between the sources, yield consistent values for the relevant observables when measured in physical units. 
For this purpose, we analyzed the full electric field and its nonperturbative component at the midpoint between the sources for various distances:  $d = 0.723 - 0.760$  fm using six different lattice setups,  $d = 0.855 - 0.959$  fm using six different setups, and  $d = 1.013 - 1.060$  fm using four setups (see Fig.~\ref{fig:scaling}).

We extract the nonperturbative part of the longitudinal electric field, $E_x^{\text{NP}}$, by systematic application of the {\em curl method} (details in Ref.~\cite{Baker:2019gsi}).
In Fig.~\ref{fig:Ex_3d} we present two 3D plots giving the space distribution of the full longitudinal electric field $E_x$ (left panel) and of its nonperturbative part (right panel): in the latter, a tube-like shape clearly emerges, fairly uniform in the longitudinal direction, up to a small asymmetry (the field at $x_l=0$ is a bit higher than the field at $x_l=d$) probably due to the asymmetric structure of the adopted lattice operator (see Fig.~\ref{fig:op_W}), which the smearing procedure is unable to balance. 
\begin {figure*}[htb]
\centering
\includegraphics[width=0.55\linewidth,clip]{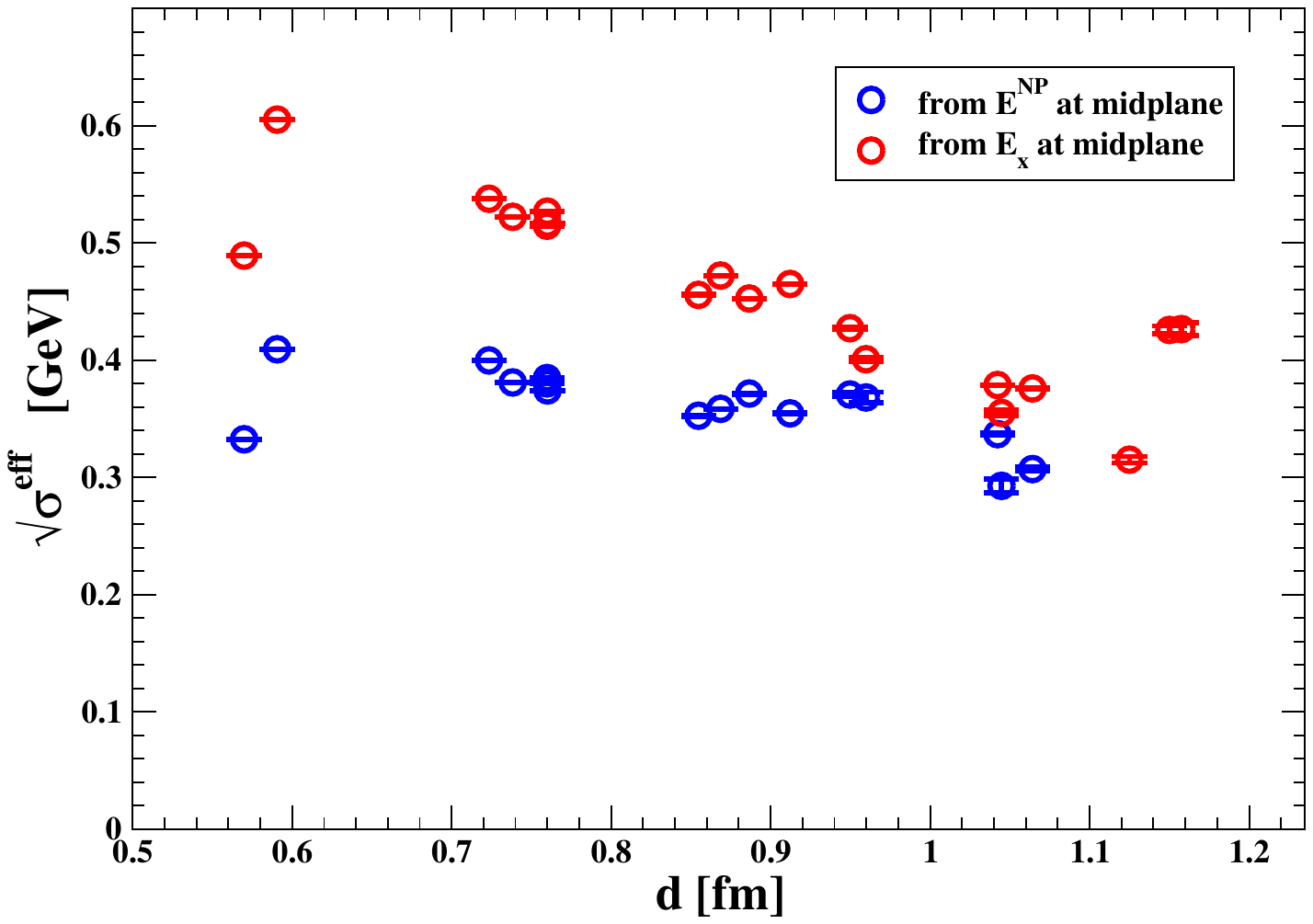}
\hspace{0.2cm}
\includegraphics[width=0.55\linewidth,clip]{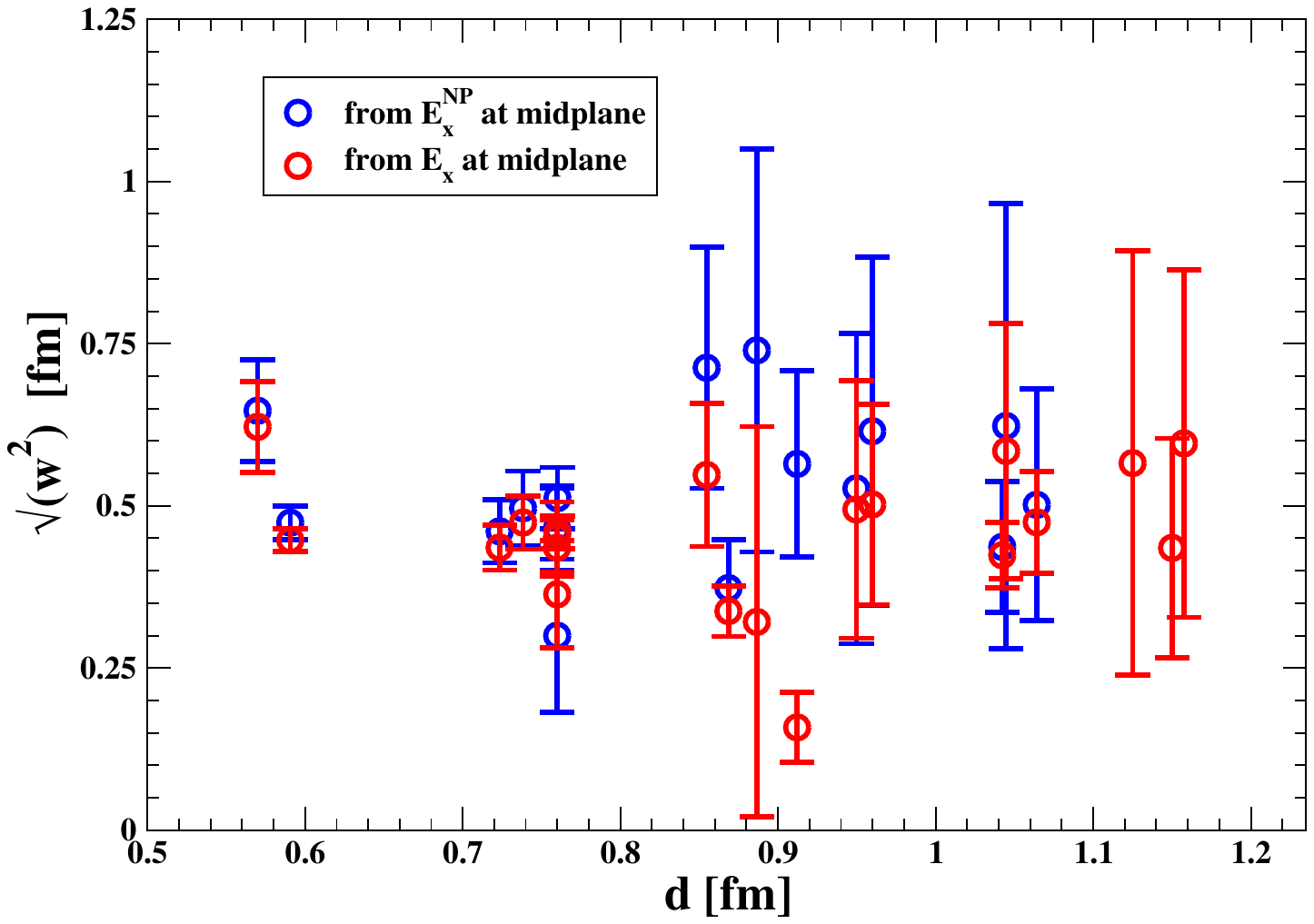}
\caption{ Behaviour of the effective string tension $\sigma_{\rm eff}$ (top)  and of the flux tube width $w$ (bottom) 
with the distance $d$ between the sources, for the full longitudinal electric field (red circles) and its nonperturbative part (blue circles).}               
\label{fig:string}
\end{figure*}

\begin {figure*}[thb]
\centering
\includegraphics[width=0.55\linewidth,clip]{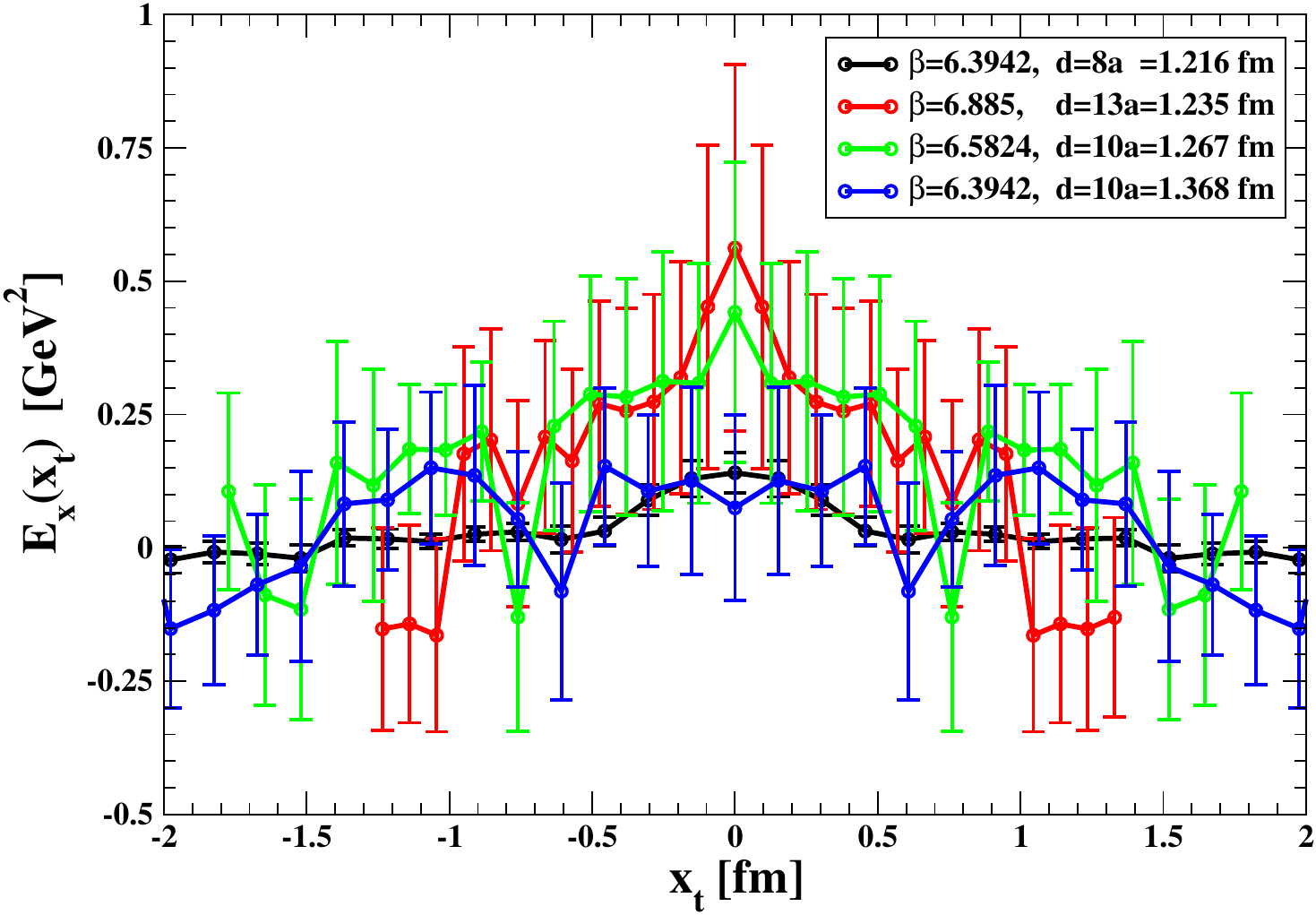}
\includegraphics[width=0.55\linewidth,clip]{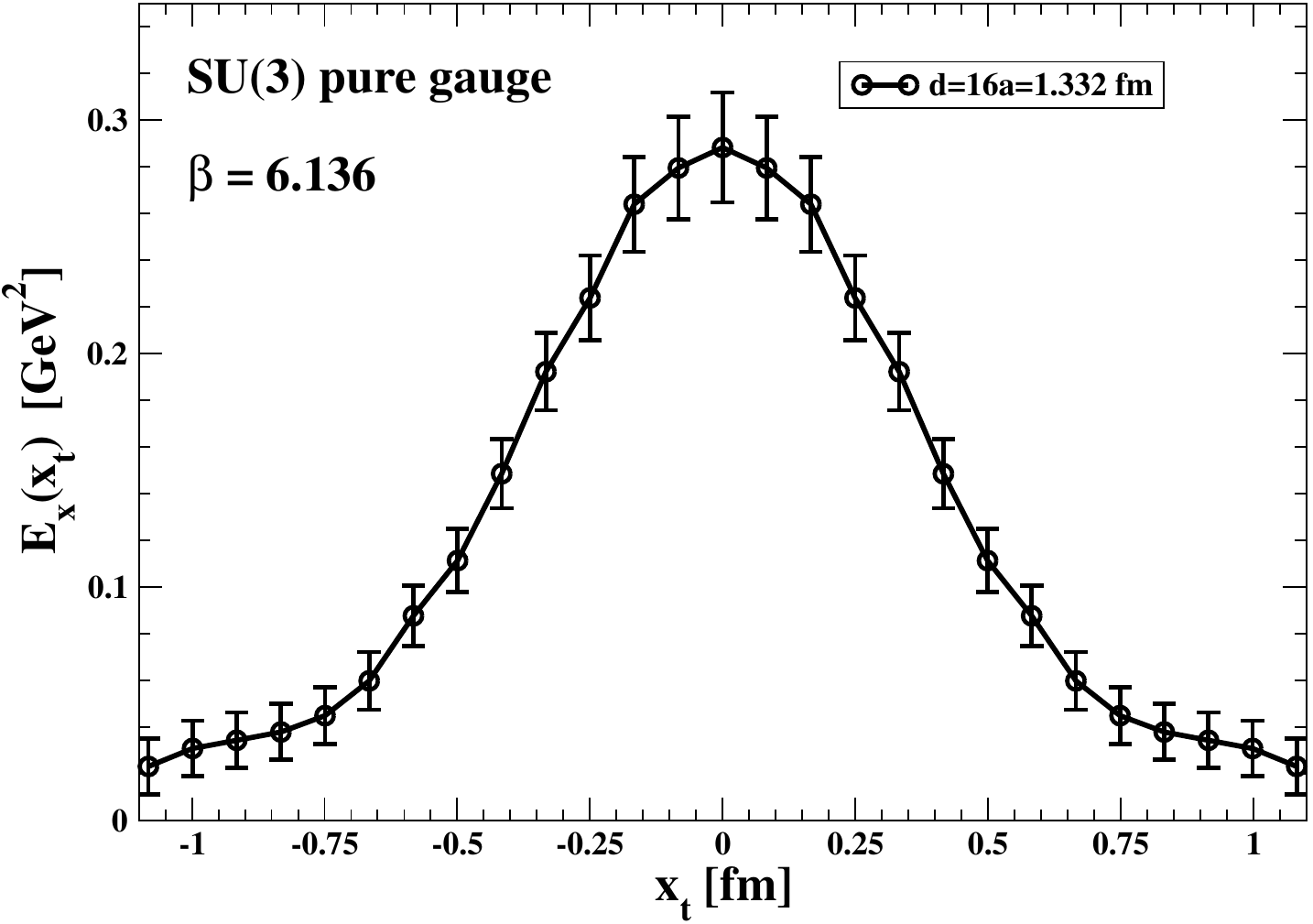}
\caption{Transverse profiles of the full longitudinal electric field on the midplane, for various distances between the sources in full QCD (top). Transverse profile of the full longitudinal electric field on the midplane in the case of pure SU(3) gauge theory at physical distance $d=1.332$ fm between the sources (bottom).}       
\label{fig:stringbreaking}
\end{figure*}

To quantitatively characterize the shape and specific properties of the flux tube formed by the longitudinal electric field, we numerically computed two expressions involving the following integrals:
\begin{equation}
\sigma_\mathrm{eff} = \int d^2 x_t \ \frac{(E_x^\mathrm{NP}(x_t))^2}{2} \ , 
\label{sigma_eff}
\end{equation}
\begin{equation} 
w = \sqrt{\frac{\int d^2x_t \, x_t^2 E_x^\mathrm{NP}(x_t)}{\int d^2x_t \, E_x^\mathrm{NP}(x_t)}} \ ,
\label{width}
\end{equation}
with the longitudinal electric field taken at the midplane between the sources.
The first of them represents a quantity which has the dimension of an energy per unit length, similarly to the string tension. The expression in Eq.~(\ref{width}) gives an estimate of the width of the flux tube.
The integrals in Eqs.(\ref{sigma_eff}) and (\ref{width}) are numerically computed using the trapezoidal rule. 
These calculations were carried out for both the nonperturbative component of the longitudinal electric field and the full field.

In Fig.~\ref{fig:string}, left panel,  we compare the behavior of $\sigma_{\rm eff}$ with the 
distance $d$ between the sources for the full longitudinal electric field
and its nonperturbative part: while for the full field $\sigma_{\rm eff}$ tends to decrease, for the nonperturbative part it is fairly stable. 
This difference in behavior is expected: the full field on the midplane also includes the perturbative contribution, which becomes increasingly insignificant as the distance between the sources grows.

In Fig.~\ref{fig:string}, right panel,  we show a similar comparison for the width. 
Unlike the case of $\sigma_{\rm eff}$, the flux tube width remains stable over a broad range of distances, despite uncertainties that increase with $d$. Within these uncertainties, the width is generally consistent between the full field and the nonperturbative component. The uncertainties in estimating the width are substantially larger than those for the string tension, due to the width being defined as a ratio of two numerical integrals.

From the constant fit of the data for string tension and string width we get $\sqrt{\sigma_{\rm eff}} \approx 0.4\ \mathrm{GeV}$, and 
$w \approx 0.5\ \mathrm{fm}$.

\section{Possible evidence for string breaking}
In the presence of light quarks, the string connecting a static quark-antiquark pair is expected to break at large distances due to the creation of a light quark-antiquark pair, 
which can then recombine with the static quarks to form two static-light mesons. 
Previous estimates of the string-breaking distance have been obtained by identifying the point where the Wilson loop and the static-light meson operator exhibit equal overlap with the ground state.
In Ref.~\cite{Bali:2005fu}, a study of  $N_f = 2$  lattice QCD with a pion mass around 640 MeV found a string-breaking distance of  $d^* \simeq 1.248(13) \, \text{fm}$. 
A more recent analysis~\cite{Koch:2018puh} with  $N_f = 2+1$  flavors, using nonperturbatively improved dynamical Wilson fermions with pion and kaon masses of approximately 280 MeV and 460 MeV, respectively, reported a string-breaking distance of  $d^* \approx 1.216 \, \text{fm}$.
In this paper, we directly examine the nonperturbative gauge-invariant longitudinal electric field, $\vec{E}^\text{NP}$, in the region between two static sources, which is responsible for forming a well-defined flux tube characterized by a nonzero effective string tension, $\sigma_{\rm eff}$, and width, $w$. To illustrate this, in Fig.~\ref{fig:stringbreaking} (left panel), we show the transverse profile of the complete longitudinal electric field on the midplane between two sources positioned at distances of 1.216 fm, 1.235 fm, 1.267 fm, and 1.368 fm. At the farthest distance, the signal is almost completely.
One might argue that this is due to signal-to-noise degradation as the distance increases in lattice units. 
However, this interpretation is contradicted by the presence of a clear signal at even larger lattice-unit distances, yet smaller physical separations, as demonstrated by the data for $d=1.235$ fm, corresponding to 13 lattice spacings in Fig.~\ref{fig:stringbreaking} (left panel).

Although we observed evidence of a nonzero total longitudinal electric field on the midplane between two sources for distances $1.216 \, \text{fm} \lesssim d < 1.368 \, \text{fm}$, 
we found no indications of a substantial nonperturbative longitudinal electric field. This prompted us to extend the analysis to shorter distances: $d \approx 1.140 \, \text{fm}, 1.157 \, \text{fm}$, and 1.183  fm. 
Even in these cases, after subtracting the perturbative component from the total longitudinal electric field, we found no clear evidence of a nonzero nonperturbative electric field. 
In other words, for $d \gtrsim 1.140 \, \text{fm}$, there is no indication of flux tube formation between the two static color sources

We also observed that when the physical quark separation $d$ exceeds a certain threshold $d^*$, 
our method for extracting the longitudinal nonperturbative electric field, $E_x^{\rm NP}$, becomes ineffective due to large uncertainties. 
Given that a well-defined nonperturbative flux tube exists up to approximately $d \simeq 1.064 \, {\text{fm}}$ and that 
we cannot reliably estimate the nonperturbative field beyond $d \simeq 1.140$ fm, we approximate this threshold as $1.064 \, \text{fm} \lesssim d^\ast \lesssim 1.140 \, \text{fm}$.
Note that our estimate is somewhat smaller than that of Ref.~\cite{Koch:2018puh}, as we are considering QCD with (2+1) dynamical fermions at physical masses. 
Additionally, our determination of the string-breaking distance is slightly lower than the recent estimate extrapolated to physical light-quark masses 
in Ref.~\cite{Bulava:2024jpj}, but it aligns well with the extrapolated string-breaking distance for real QCD found in Ref.~\cite{Bali:2005fu}.
Additional numerical evidence suggests that above $d^*$, no improvement is observed even if the lattice spacing is reduced while keeping $d^*$ fixed. 
In contrast, in SU(3) pure gauge theory—where, by definition, the string remains unbroken—a clear signal persists at physical distances around $d \sim d^*$ (see Fig.~\ref{fig:stringbreaking}, right panel).

In our approach, the string breaking distance $d^\ast$ should appear as a drop in the nonperturbative electric field and, 
consequently, a decrease in the effective string tension, $\sqrt{\sigma_\mathrm{eff}}$. 
Although our numerical data does not show a definitive drop in $\sqrt{\sigma_\mathrm{eff}}$ the fact that our method fails beyond $d^\ast$ in 2+1 flavor QCD—but not in pure gauge theory—may itself indicate string breaking.

Based on this analysis, we cannot make a definitive claim about the onset of string breaking; instead, we provide some indirect hints of its occurrence, which require further testing to confirm.

\section{Conclusions}

We investigated the behavior of the nonperturbative, gauge-invariant longitudinal electric field, $\vec{E}^\text{NP}$, 
between a static quark and antiquark through Monte Carlo simulations of QCD with (2+1) dynamical staggered fermions at physical masses.
Our simulations spanned a range of coupling values within the continuum scaling regime and included source separations from approximately 0.5 fm up to about 1.37 fm. 

After subtracting the perturbative component, the longitudinal electric field forms a flux tube whenever the source separation is below $d^* \simeq 1.1 \, {\text{fm}}$. 
This flux tube can be described by two quantities:  $\sigma_{\rm eff}$, related to the effective string tension, and the width $w$, both determined by numerically integrating our lattice data on the midplane between sources.

Beyond the distance $d^*$, the longitudinal nonperturbative field is generally consistent with zero, within large numerical uncertainties. 
Our approach does not allow us to determine whether this vanishing signal at larger distances is due to signal-to-noise degradation or to the onset of string breaking. 
We have provided some numerical arguments in favor of string breaking, but further investigation is needed to confirm this interpretation.

\section*{Acknowledgements}
This investigation was in part based on the MILC collaboration's public lattice gauge theory code (\url{https://github.com/milc-qcd/}). 
Numerical calculations have been made possible through a CINECA-INFN agreement, providing access to HPC resources at CINECA. PC, LC and AP acknowledge support from INFN/NPQCD project. 
VC acknowledges support by  the Deutsche Forschungsgemeinschaft  (DFG, German Research Foundation) through the CRC-TR 211  ``Strong-interaction matter under extreme conditions'' -- project number 315477589 -- TRR 211. This work is (partially) supported by ICSC – Centro Nazionale di Ricerca in High Performance Computing, Big Data and Quantum Computing, funded by European Union – NextGenerationEU.


\providecommand{\href}[2]{#2}\begingroup\raggedright\endgroup

\end{document}